\documentclass[prd,twocolumn,showpacs,letterpaper,nofootinbib]{revtex4}
\usepackage{graphics}
\usepackage{epsfig} 
\usepackage{amsmath}
\usepackage{multirow}

\input epsf

\newcommand{\be}{\begin{equation}}
\newcommand{\ee}{\end{equation}}
\newcommand{\bea}{\begin{eqnarray}}
\newcommand{\eea}{\end{eqnarray}}
\def\pslash{{\cal P}{\hbox{\kern-6pt $\slash$}}}

\long\def\comment#1{}
\begin{document}
\title{Dark sectors of the Universe: A Euclid survey approach}
\author{Isaac Tutusaus}
\email{isaac.tutusaus@irap.omp.eu}
\affiliation{Universit\'e de Toulouse, UPS-OMP, IRAP and CNRS, IRAP, 14, avenue Edouard Belin, F-31400 Toulouse, France}
\author{Brahim Lamine}
\email{brahim.lamine@irap.omp.eu}
\affiliation{Universit\'e de Toulouse, UPS-OMP, IRAP and CNRS, IRAP, 14, avenue Edouard Belin, F-31400 Toulouse, France}
\author{Alain Blanchard}
\email{alain.blanchard@irap.omp.eu}
\affiliation{Universit\'e de Toulouse, UPS-OMP, IRAP and CNRS, IRAP, 14, avenue Edouard Belin, F-31400 Toulouse, France}
\author{Arnaud Dupays}
\email{arnaud.dupays@irap.omp.eu}
\affiliation{Universit\'e de Toulouse, UPS-OMP, IRAP and CNRS, IRAP, 14, avenue Edouard Belin, F-31400 Toulouse, France}
\author{Yvan Rousset}
\email{alain.blanchard@irap.omp.eu}
\affiliation{Universit\'e de Toulouse, UPS-OMP, IRAP and CNRS, IRAP, 14, avenue Edouard Belin, F-31400 Toulouse, France}
\author{Yves Zolnierowski}
\email{zolniero@lapp.in2p3.fr}
\affiliation{Laboratoire d'Annecy-le-Vieux de
Physique des Particules, CNRS/IN2P3 and Universit\'{e}
Savoie Mont Blanc, 9 Chemin de Bellevue, BP 110, F-74941 Annecy-le-Vieux cedex, France}

\date{\today}

\begin{abstract}
  In this paper we study the consequences of relaxing the hypothesis
  of the pressureless nature of the dark matter component when determining constraints on dark
  energy. To this aim we consider simple generalized dark matter
  models with constant equation of state parameter. We find that
  present-day low-redshift probes (type-Ia supernovae and baryonic
  acoustic oscillations) lead to a complete
  degeneracy between the dark energy and the dark matter
  sectors. However, adding the cosmic microwave background (CMB) high-redshift
  probe restores constraints similar to those on the standard $\Lambda$CDM
  model. We then examine  the anticipated constraints from the galaxy
  clustering probe of the future
  Euclid survey on the same class of models, using a Fisher forecast estimation.
 We show that the Euclid survey allows us to break the degeneracy
 between the dark sectors, although the constraints on dark energy are
 much weaker than with standard dark matter. The use of CMB in
 combination  allows us to restore the high precision on the dark energy
 sector constraints.
\end{abstract}

\maketitle
\vskip 1cm

\section{INTRODUCTION}

The $\Lambda$CDM framework offers a simple description of the
properties of our Universe with a very small number of free
parameters. It reproduces remarkably well a wealth of high-quality
observations which allow us to determine with high precision the few
parameters describing the model (typically below 5\% in most of the
parameters~\cite{Planck2015}). Therefore, more than fifteen years
after the establishment of the accelerated expansion of the
Universe~\cite{Riess,Perlmutter}, the $\Lambda$CDM cosmological model
remains the current standard model in cosmology. However, the dark
contents of the Universe remain largely unidentified: no direct
detection of a dark matter particle has been achieved, and the
theoretical basis of the observed cosmological constant is not clearly
established, especially with respect to the issue of gravitational
effects of quantum vacuum energy (the cosmological constant problem
$-$see \cite{JMartin} for an extended review). In this context, a
large variety of explanations have been proposed beyond a simple
Einstein cosmological constant: scalar field domination known as
quintessence~\cite{quintessence}, generalized gravity theory beyond
general relativity~\cite{MG} or even inhomogeneous cosmological
models~\cite{inhomogeneous}. An extensive review on constraints on
cosmological models with the Euclid satellite can be found
in~\cite{Amendola2013}. In addition, it has already been noticed that
the pressureless (cold) nature of dark matter itself is not firmly
established~\cite{GDM}. Finally, even the separation of the dark
sector in physically independent components such as a dark matter
component and a dark energy may not be relevant with cosmological
observations alone~\cite{kunz}. 

In this paper, we examine the consequences of considering 
nonpressureless dark matter when constraining the dark energy sector,
with present-day observations and in the context of the future Euclid
survey mission. We focus on the simplest models, both for the dark
matter and the dark energy sectors. Namely, we assume constant
equation of state for both sectors: $P=w_{DM}\rho$ for the dark
matter sector ($w_{DM}\neq0$ implies that the dark matter component
has some pressure), and $P=w_{DE}\rho$ for the dark energy sector. Section~\ref{section:2}
summarizes the constraints obtained on the previous two parameters
using the present-day data, while Sec.~\ref{section:3} shows the
improvements that are expected with the Euclid survey.

\section{DARK CONTENT(S) OF THE UNIVERSE }\label{section:2}
In this section, we use observations from type-Ia supernovae (SNIa),
the baryonic acoustic oscillations (BAO) and the cosmic microwave
background (CMB) to constrain cosmological models in the presence of
variations in both the standard dark energy and the standard dark matter sectors.

\subsection{Method and data samples}\label{data}
In the following, we do not work with
the full likelihood of the previously mentioned probes,
but with compressed forms of them. They have been shown to be faster
and easier to evaluate and still remain accurate for the most common
cases. We assume a flat universe, a Robertson-Walker metric and
Friedmann-Lema\^{i}tre dynamics through all the work. The
Friedmann-Lema\^{i}tre equation is then given by
\begin{align}\label{fl}
H^2(z) = &\frac{8\pi G}{3}\left(\rho_r+\rho_b + 
\rho_{DM} +
\rho_{DE}\right)\,,
\end{align}
where we have already imposed the constraint of a flat Universe, i.e.
the sum of the density parameters ($\Omega\equiv 8\pi G\rho/(3H^2)$) is
equal to one. In this work we adopt the value given in
\cite{Komatsu2011} for the radiation density parameter,
\begin{equation}
\Omega_r=\Omega_{\gamma}(1+0.2271 N_{eff}f)\,,
\end{equation}
with $N_{eff}=3.046,\,\Omega_{\gamma}=2.469\times 10^{-5}h^{-2}$ and
$f=(1+(0.3173\cdot 187\times
10^3(m_{\nu}/94\,\text{eV})(1+z)^{-1})^{1.83})^{1/1.83}$, where $m_{\nu}$ is the
sum of the mass of three neutrino families, which we have approximated to
be 0.  We also assume that the different fluids present in
 the Universe are noninteracting with constant equation of state. The Friedmann-Lema\^{i}tre
 equation then reads
\begin{align}\label{ecdm}
\frac{H^2(z)}{H_0^2}=&\Omega_r(1+z)^4+\Omega_b(1+z)^3+\\
&(\Omega_m-\Omega_b)(1+z)^{3(1+w_{DM})}+\nonumber\\
&(1-\Omega_r-\Omega_m)(1+z)^{3(1+w_{DE})}\nonumber\,.
\end{align}

In order to determine the constraints on the parameters under study we
assume a Gaussian likelihood and therefore use a $\chi^2$ minimization
procedure; i.e. we minimize the function,
\begin{equation}\label{chisq}
\chi^2=(\textbf{u}-\textbf{u}_{data})^TC^{-1}(\textbf{u}-\textbf{u}_{data})\,,
\end{equation}
where $\textbf{u}_{data}$ is the data vector, $\textbf{u}$ is the
compressed likelihood parameters representation of the data in the models
under investigation and $C$ is the covariance matrix of the data.

As we are combining essentially independent probes, we
obtain the total $\chi^2$ function as the sum of each of them.

\subsubsection{The SNIa sample}
For the SNIa data we use the  compressed form of the likelihood of the JLA sample~\cite{Betoule2014},  which includes  740 SNIa from the full three years SDSS survey~\cite{Sako2014} and from the compilation assembled in \cite{Conley2011}, comprising SNIa from SNLS, HST and several nearby experiments. The $\chi^2$ for the SNIa probe then reads
\begin{equation}
\chi^2=\textbf{r}C_b^{-1}\textbf{r}^T\,,
\end{equation}
with
\begin{equation}
\textbf{r}=\boldsymbol{\mu}_b-M-5\log_{10}D_L(\boldsymbol{z}_b)\,,
\end{equation}
where $\boldsymbol{\mu}_b$ is the vector of distance modulus at
redshift $\boldsymbol{z}_b$ (binned redshifts), $M$ is a free
normalization parameter, $C_b$ is the covariance matrix of
$\boldsymbol{\mu}_b$ and $D_L$ is the luminosity distance (see
\cite{Betoule2014} for detailed explanations).
It is important to note that the normalization parameter $M$ must be
left free in the fit and marginalized over when deriving
uncertainties, in order to avoid introducing artificial constraints on
the Hubble constant $H_0$.

\subsubsection{Baryonic acoustic oscillations}
Present-day BAO analyses are usually performed through a spherical average of their scale measurements, resulting in a constraint on  a combination of the angular scale and redshift separation. It is commonly expressed as a measure on
\begin{equation}
d_z=\frac{r_s(z_{drag})}{D_v(z)}\,,
\end{equation}
with
\begin{equation}
D_v(z)=\left((1+z)^2D_A^2\frac{cz}{H(z)}\right)^{1/3}\,,
\end{equation}
where $D_A$ is the angular-diameter distance, $H(z)$ is the Hubble
parameter and $z_{drag}$ is the redshift at which the baryons are
released from the Compton drag of the photons. The comoving scale
$r_s(z_{drag})$ corresponds to the total distance a sound wave can
travel prior to $z_{drag}$ and it can
be expressed in terms of an integral over the sound
velocity,
\begin{equation}
r_s(z_{drag})=\int_{z_{drag}}^{\infty}\frac{c_s(z)\,\text{d}z}{H(z)}\,,
\end{equation}
where
\begin{equation}
c_s(z)=\frac{c}{\sqrt{3(1+R_b(z))}},\hspace{15pt} R_b(z)=\frac{3\rho_b}{4\rho_{\gamma}}\,,
\end{equation}
with $\rho_b$ being the baryon density and $\rho_\gamma$ the photon density. 
For this work, we use the BAO measurements at $z= 0.106$~\cite{Beutler2011}, $z=0.35$~\cite{Pad2012} and $z=0.57$~\cite{Anderson2012}, following Planck
Collaboration XVI~\cite{PlanckXVI2013}, which consists in the data vector $d_z=(0.336,0.1126,0.07315)$ and the
inverse of the covariance matrix
$C^{-1}=\text{diag}(4444,215156,721487)$. We use the fitting formulas
from \cite{EHu1998} to compute $z_{drag}$ and the value they provide
for $R_b(z)=3.15\times 10^4\omega_b \Theta_{2.7}^{-4}(1+z)^{-1}$, with
$\Theta_{2.7}=T_{CMB}/2.7\text{K}=2.728/2.7$ and $\omega_b\equiv
\Omega_b h^2$.

\subsubsection{Cosmic microwave background}
As shown in \cite{WM2007}, a significant part of the the information coming from the CMB can be
compacted into a few numbers, the so-called reduced parameters: the
scaled distance to recombination $R$, the angular scale of the sound
horizon at recombination $\ell_a$ and the reduced density parameter of baryons $\omega_b$. For a flat universe their expressions are
given by
\begin{align}
&R\equiv \sqrt{\Omega_m H_0^2}\int_0^{z_{CMB}}\frac{\text{d}z}{H(z)},\\
 \ell_a\equiv &\frac{\pi c}{r_s(z_{CMB})}\int_0^{z_{CMB}}\frac{\text{d}z}{H(z)},\hspace{25pt}\omega_b\equiv \Omega_b h^2\,,\nonumber
\end{align}
where we use the fitting formulas from \cite{HS1996} to compute
$z_{CMB}$, the redshift of the last scattering epoch. We use the data obtained from the Planck~2015 data release~\cite{Planck2015XIV}, where the compressed likelihood parameters are
obtained from the Planck temperature-temperature plus the low-$\ell$
Planck temperature-polarization likelihoods. We use the
compressed likelihood parameters obtained when marginalizing over the
amplitude of the lensing power spectrum for the lower values, which
leads to a more conservative approach: $(\ell_a,R,\omega_b)=(301.63\pm
0.15, 1.7382\pm 0.0088,0.02262\pm 0.00029)$ with the normalized
covariance matrix,
\begin{equation}
C=\left(
\begin{array}{ccc}
1.0000 & 0.64 & -0.55\\
0.64 & 1.0000 & -0.75\\
-0.55 & -0.75 & 1.0000
\end{array}\right)\,.
\end{equation}

\subsection{Models}\label{section:models}
In this section we present  different cosmological models generalizing the standard $\Lambda$CDM model. Constraints on cosmological parameters for these models, obtained with the existing data presented in Sec.\,\ref{data}, are provided in the following section.

\subsubsection{$w$CDM model}
We first study a reference model with standard cold dark matter and a
dark energy component with constant equation of state parameter:
$w_{DM}=0$ and $w_{DE}=w$ in Eq.\,(\ref{ecdm}). We denote this model $w$CDM (see for example \cite{Cheng} for a previous study of this model).

\subsubsection{$\epsilon$CDM model}
We define the $\epsilon$CDM model by assigning $w_{DM}=0+\epsilon$ and
$w_{DE}=-1$. This is the simplest version of the generalized dark
matter (GDM) model~\cite{GDM}, where the cold dark matter is
replaced by a GDM but keeping the speed of sound and viscosity equal
to 0 (see \cite{Thomas2016} for a detailed study of this model and
other GDM models). Since in this $\epsilon$CDM model we are modifying
the matter component in the Universe and it has an extremely important
role in the CMB era we must adapt the computation of $z_{CMB}$ and
$z_{drag}$ by changing $(\Omega_m-\Omega_b)$ by
$(\Omega_m-\Omega_b)(1+z_{CMB})^{3\epsilon}\approx
(\Omega_m-\Omega_b)(10^3)^{3\epsilon}$ and compute $R$ as
$\sqrt{(\Omega_b+(\Omega_m-\Omega_b)(1+z_{CMB})^{3\epsilon})H_0^2}\int_0^{z_{CMB}}\,\text{d}z/H(z)$.
This comes from the fact that we change $\Omega_{DM}(z)\equiv
\Omega_m(z)-\Omega_b(z)=\Omega_{DM}(1+z)^3$ by
$\Omega_{DM}(z)=\Omega_{DM}(1+z)^{3(1+\epsilon)}$; therefore, the
effective $\Omega_{DM}$ that should appear in the CMB era is given by

\begin{equation}
\Omega_{DM}^{eff}=\Omega_{DM}\frac{(1+z_{CMB})^{3(1+\epsilon)}}{(1+z_{CMB})^3}=\Omega_{DM}(1+z_{CMB})^{3\epsilon}\,.
\end{equation}

\subsubsection{$\epsilon w$CDM model}
Finally, we consider an extended version of the $\epsilon$CDM model
allowing for variations in the dark matter and the dark
energy sectors at the same time, with a constant dark energy
equation of state parameter different from $-1$. We denote such a
model as the $\epsilon w$CDM model, having two parameters for the dark sector, $w_{DM}=\epsilon$ and $w_{DE}=w$. Notice that for this model we must
keep the previous modifications on $z_{CMB}$, $z_{drag}$ and $R$ since
we modify the matter component.

\subsection{Results}\label{section:results}
In order to perform the $\chi^2$ minimization described in Sec.\,\ref{data}, we use the MIGRAD application from the MINUIT tool,
conceived to find the minimum value of a multiparameter function and
analyze the shape of the function around the minimum \cite{minuit}. To
compute the contours and the errors on the parameters we use a
statistical method based on profile likelihoods. For a given parameter
$\theta$ we fix it at different values and, for each of them, we
minimize the $\chi^2$ function with respect to all the remaining
parameters. For each fixed value of $\theta$ we keep the
$\chi^2_{min}$ value. The curve interpolated through
$\{\theta(i),\chi^2_{min}(i)\}$ points and offset to 0 is the
so-called $\theta$ profile likelihood $\Delta \chi^2(\theta)$ (see
\cite{Couchot,PlanckPF} for a more detailed description and a
comparison with the common Markov chain Monte Carlo method).

\begin{figure*}[t]
\includegraphics[scale=0.45]{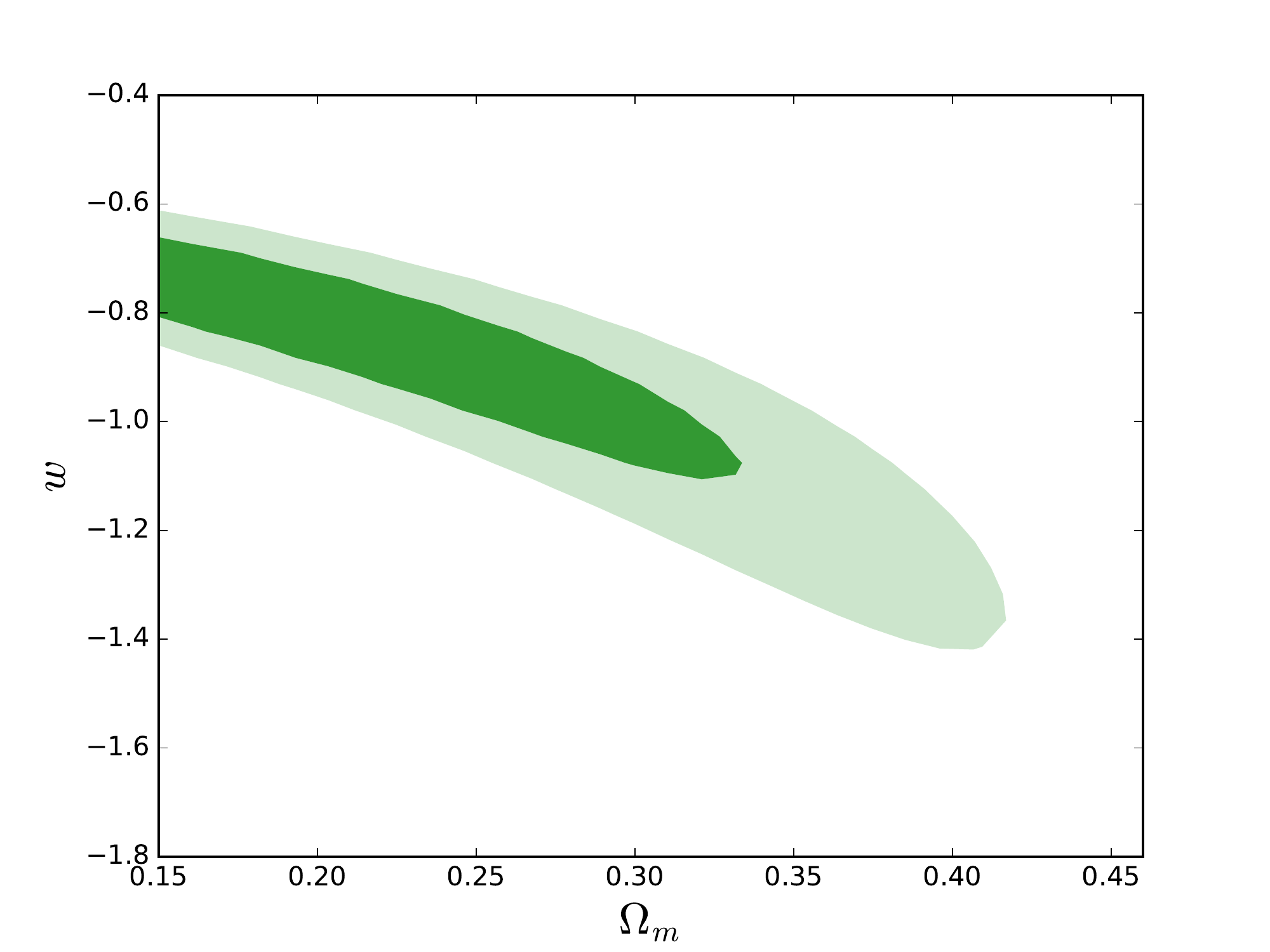}\includegraphics[scale=0.45]{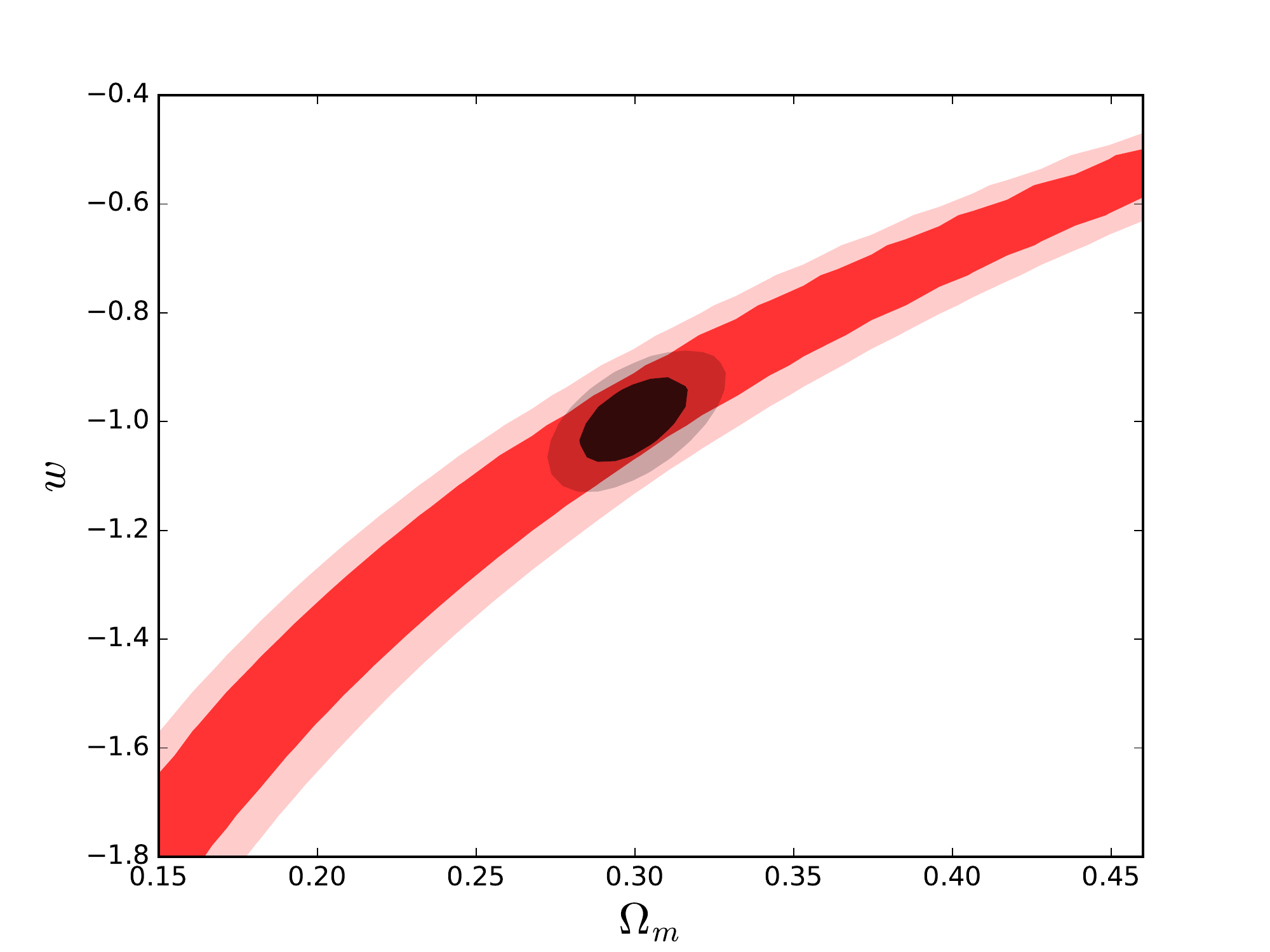}
\caption{Confidence contours at 68\% and 95\% ($\Delta \chi^2=2.30$
  and $\Delta \chi^2=6.17$, respectively) for the $\Omega_m$ and $w$
  cosmological parameters of the $w$CDM model. Left panel: contours
  obtained using the SNIa and the BAO cosmological probes with
  $\omega_b=0.02262$ fixed. Right panel: the red (large) contours correspond
  to the CMB probe while the black (small) contours account for the
  combination of the three probes: SNIa, BAO and CMB.}\label{fig1}
\end{figure*}

In Fig.\,\ref{fig1} we show the result of our analysis for the  $w$CDM model, 
with  
the 1$\sigma$ and 2$\sigma$
contours obtained for the $\Omega_m$ and $w$ cosmological parameters.
In the left panel only the information coming
from the SNIa and the BAO probes has been used (fixing the reduced
baryon density parameter to $\omega_b=0.02262$~\cite{Planck2015XIV}), while in the right
panel the red contours correspond to the CMB probe and the black ones
correspond to the combination of the three probes: SNIa+BAO+CMB. In
these cases we have not fixed the baryon density as this quantity is 
well constrained by the CMB probe. We
have marginalized over $H_0$ in both panels. The constraints we have obtained for the different models 
are summarized in Table\,\ref{table1}. For the $w$CDM model, our constraints are 
(logically) very similar to those of \cite{Betoule2014},  whose authors used the
BAO, SNIa and CMB through temperature fluctuations from Planck~2013
and polarization fluctuations from WMAP.

\begin{figure*}[t]
\includegraphics[scale=0.45]{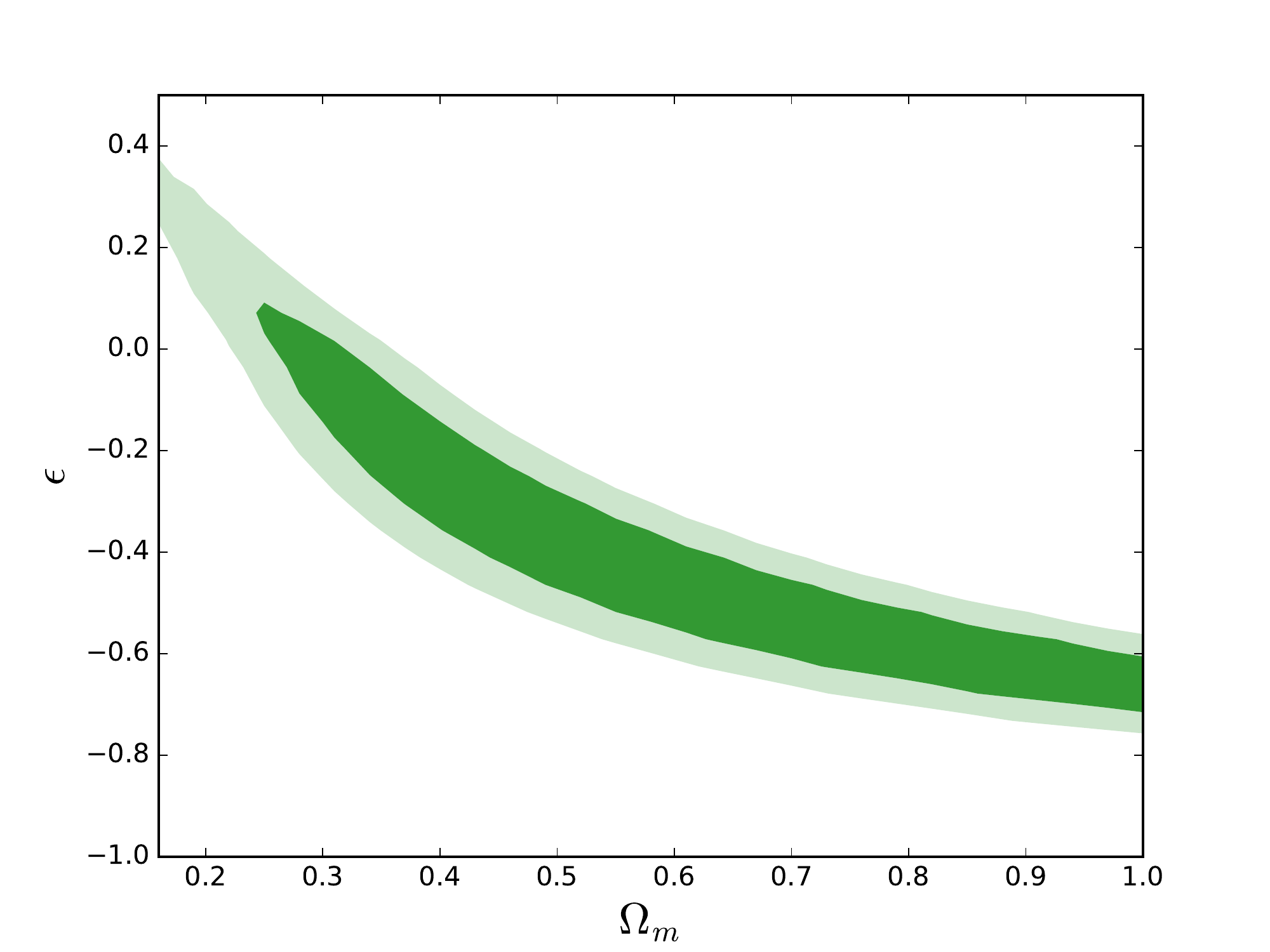}\includegraphics[scale=0.45]{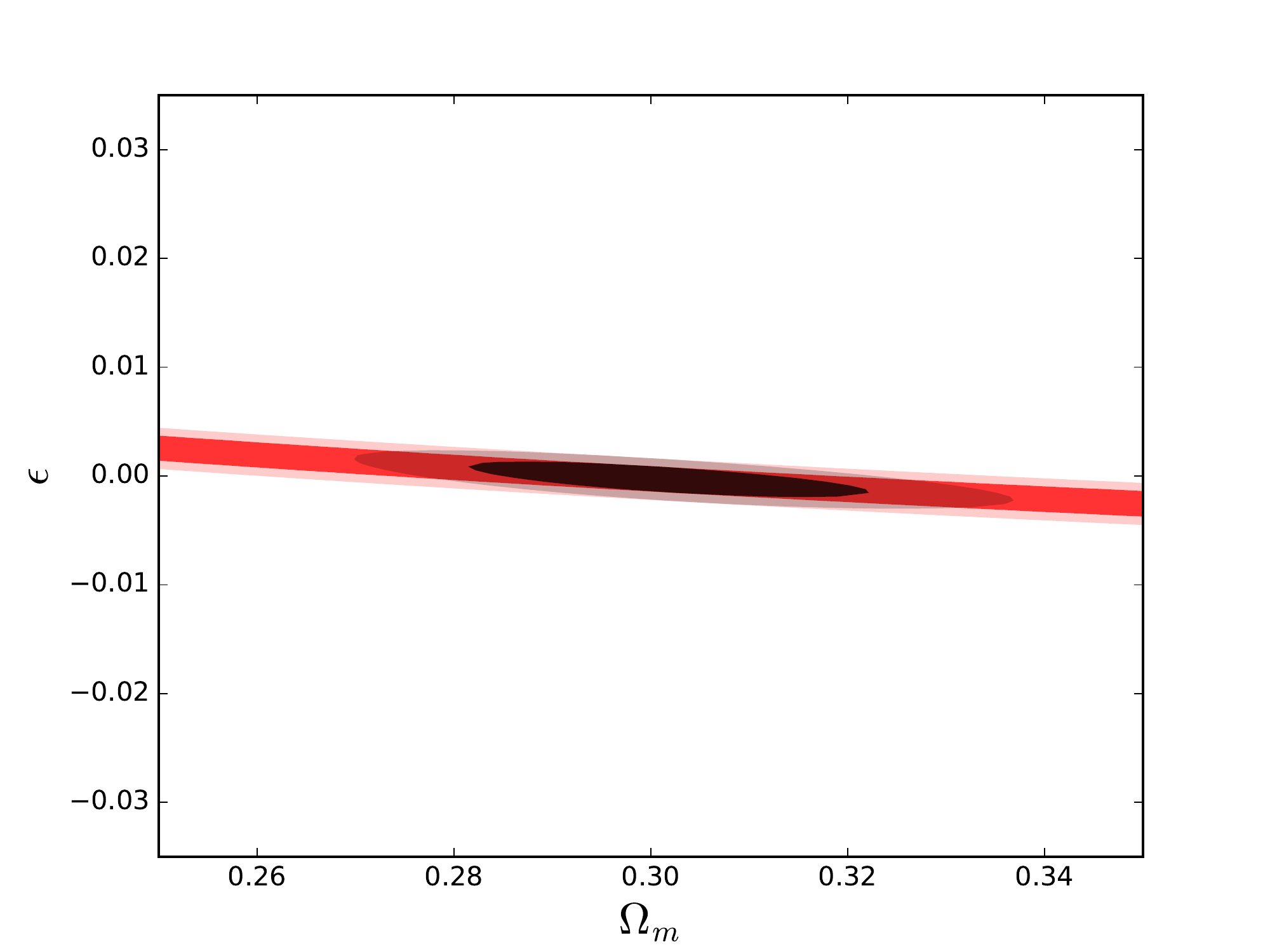}
\caption{Confidence contours at 68\% and 95\% ($\Delta \chi^2=2.30$
  and $\Delta \chi^2=6.17$, respectively) for the $\Omega_m$ and $\epsilon$
  cosmological parameters of the $\epsilon$CDM model. Left panel: contours
  obtained using the SNIa and the BAO cosmological probes with
  $\omega_b=0.02262$ fixed. Right panel: the red (large) contours correspond
  to the CMB probe while the black (small) contours account for the
  combination of the three probes: SNIa, BAO and CMB.}\label{fig2}
\end{figure*}

In Fig.\,\ref{fig2} the 1$\sigma$ and 2$\sigma$ contours for the
$\Omega_m$ and $\epsilon$ parameters of the $\epsilon$CDM model are
represented. As in Fig.\,\ref{fig1}, the left panel corresponds to the
result using only SNIa+BAO (fixing $\omega_b=0.02262$), while the
right panel shows the CMB contours and the combination of the three
probes, with marginalization over the baryon density. We have
marginalized over $H_0$ in all cases. The specific constraints we have
obtained are $\Omega_m=0.301^{+0.014}_{-0.013}$ and
$\epsilon=-0.0003\pm 0.0011$ (errors at 1$\sigma$ on one parameter),
which clearly differ from the result in \cite{Avelino2012}, where they
provide $\epsilon=0.009\pm 0.002$ at 3$\sigma$ confidence level. This
difference can be due to the use of different cosmological
probes. However, our results are compatible with \cite{Thomas2016}
where the authors provide $\epsilon=0.00063^{+0.00108}_{-0.00112}$ at
2$\sigma$ confidence level, using Planck data plus lensing
information and the BAO probe. 

\begin{figure*}[t]
\includegraphics[scale=0.45]{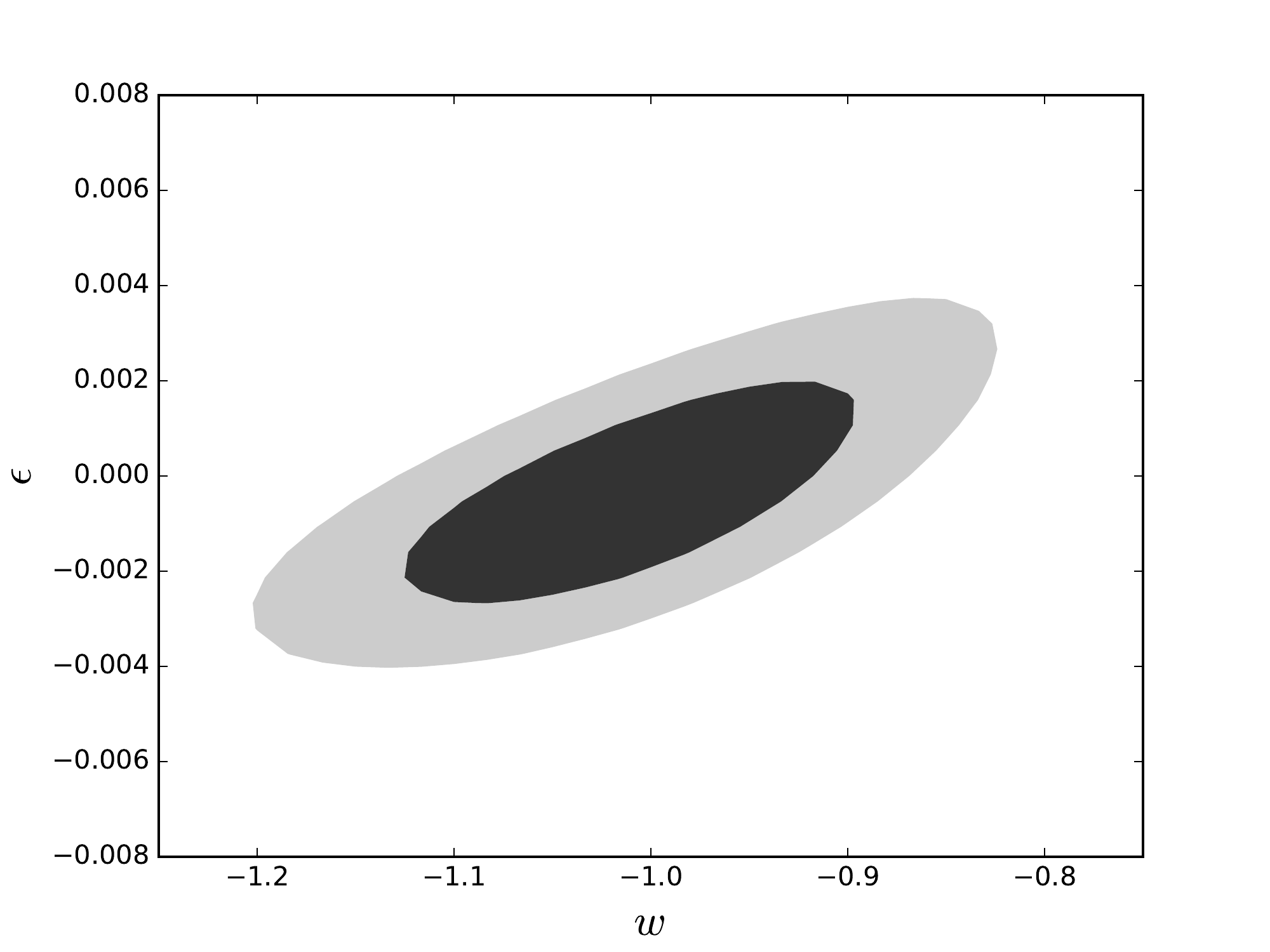}
\caption{Confidence contours at 68\% and 95\% ($\Delta \chi^2=2.30$
  and $\Delta \chi^2=6.17$, respectively) for the $w$ and $\epsilon$ cosmological parameters of the $\epsilon w$CDM model. The
  combination of the three probes SNIa, BAO and CMB has been used.}\label{fig3}
\end{figure*}

In Fig.\,\ref{fig3} the 1$\sigma$ and 2$\sigma$
contours for the $w$ and $\epsilon$ cosmological parameters of the
$\epsilon w$CDM model are provided. In this case only the combination of the three
probes is represented, since the contours coming only from SNIa+BAO or
only from the CMB are highly degenerated. We have marginalized
over $H_0$ and $\omega_b$. The specific obtained constraints are
$w=-1.010^{+0.075}_{-0.077}$ and $\epsilon=-0.0004\pm 0.0016$ (errors at
1$\sigma$ on one parameter), which are slightly worse than for the
$w$CDM and $\epsilon$CDM models due to the introduction of a new
degree of freedom.

All the constraints we have obtained are compatible with the standard
$\Lambda$CDM model. However, it is important to stress two points
here: first of all, we have seen the strong role of the CMB probe
(SNIa+BAO alone cannot provide us with good constraints for the
cosmological parameters) and, secondly, we have seen that the
constraints on dark matter and dark energy are not completely
independent (see Fig.\,\ref{fig3}). This implies that all the
assumptions made in one of the sectors may influence the constraints
obtained in the other one.

\section{DARK CONTENT(S) OF THE UNIVERSE: A EUCLID
  FORECAST}\label{section:3}

In this section we derive the expected precision achievable on the
previous models, using the galaxy power spectrum Euclid survey in the
linear regime.

\subsection{Method}\label{method_forecast}

In the following, the forecast is based on a Fisher matrix formalism in
a parametrized cosmological model considering the Hubble parameter
and the angular-diameter distance as observables. We rely on the
following matter power spectrum~\cite{Amendola2013}:
\begin{align}
P_{\text{matter}}(k,z)=&\frac{8\pi^2 c^4 k_0 \Delta_R^2(k_0)}{25H_0^4\Omega_m^2}T^2(k)\cdot\\
&\cdot\left[\frac{G(z)}{G(z=0)}\right]^2\left(\frac{k}{k_0}\right)^{n_s}\,,\nonumber
\end{align}
where $G(z)$ is the growth function~\cite{Carroll1992,Eisenstein1997},
$T(k)$ is the transfer function~\cite{HS1996,Eisenstein1999,EHu1998},
$k_0=0.005$ Mpc$^{-1}$, $\Delta_R^2(k_0)=2.22\times
10^{-9}$~\cite{Amendola2013} and $n_s=0.96$~\cite{Planck2015} is the
spectral index. The observed galaxy power spectrum is
different from the matter power spectrum because of the biasing of
galaxies and their velocity field. It can be related to
$P_{\text{matter}}$ by~\cite{Seo2003}
\begin{equation}
P_{\text{obs}}(k_{\perp},k_{\parallel},z)=\frac{D_A(z)^2_{\text{ref}}H(z)}{D_A(z)^2
H(z)_{\text{ref}}}P_g(k_{\perp},k_{\parallel},z)+P_{\text{shot}}\,,
\end{equation}
with
\begin{equation}
P_g(k_{\perp},k_{\parallel},z)=b(z)^2\left[1+\beta(z)\frac{k_{\parallel}^2}{k_{\parallel}^2+
k_{\perp}^2}\right]P_{\text{matter}}(k,z)\,,
\end{equation}
where $b(z)$ is the bias factor between the galaxy and matter
distributions, $\beta(z)=f_g(z)/b(z)\approx
\Omega_m^{0.6}/b(z)$~\cite{Seo2007} and $k_{\parallel},\,k_{\perp}$
stand for the parallel and transverse components of
$\textbf{k}$. $P_{\text{shot}}$ is an unknown residual noise which we neglect, since it is expected to introduce negligible
error~\cite{Samushia2011}. The ref subscript stands for the
reference cosmology.

For a given redshift interval, the Fisher matrix is given
by~\cite{Tegmark1997}
\begin{align}\label{fisher}
  F_{ij}=&\int_{-1}^1\int_{k_{\text{min}}}^{k_{\text{max}}}\frac{\partial 
\ln P_{\text{obs}}(k,\mu)}{\partial p_i}\frac{\partial \ln P_{\text{obs}}(k,\mu)}{\partial p_j}\cdot\\
  &\cdot V_{\text{eff}}(k,\mu)\frac{2\pi
    k^2\,\text{d}k\,\text{d}\mu}{2(2\pi)^3}\,,\nonumber
\end{align}
where we have changed $k_{\parallel},\,k_{\perp}$ by $k$ and $\mu$
(the modulus of \textbf{k} and the cosine of the angle between
\textbf{k} and the line of sight, respectively). According to
\cite{Seo2003} the maximum scale of the survey $k_{\text{min}}^{-1}$
has almost no effect on the results; therefore we adopt
$k_{\text{min}}=0$. The minimum scale $k_{\text{max}}^{-1}$ is used to
exclude scales affected by the nonlinear regime, where our linear
power spectra would be inaccurate. We interpolate the values given in
\cite{Seo2003} for $k_{\text{max}}$. The effective volume of the
survey is given by
\begin{align}
V_{\text{eff}}(k,\mu)&=\int\left[\frac{n(\textbf{r})P(k,\mu)}{n(\textbf{r})
P(k,\mu)+1}\right]^2\,\text{d}\textbf{r}\\
&=\left[\frac{nP(k,\mu)}{nP(k,\mu)+1}\right]^2V_{\text{survey}}\,.\nonumber
\end{align}
The last equality holds for a uniform density of galaxies. The
comoving volume of the survey $V_{\text{survey}}$ is given
by~\cite{Hogg1999}
\begin{equation}
V_{\text{survey}}=\int\text{d}\Omega\int_{z_{\text{min}}}^{z_{\text{max}}}
\text{d}z\,\frac{c(1+z)^2D_A^2(z)}{H(z)}\,,
\end{equation}
where $\Omega$ is the solid angle. According to~\cite{Seo2003} we multiply
the integrand of the Fisher matrix by an exponential suppression
$e^{-k^2\mu^2\sigma_r^2}$, with $\sigma_r=c\sigma_z/H(z)$, in order to
take into account the redshift error $\sigma_z$ of the galaxy
survey. Once we have obtained the Fisher matrix for the observables
$H(z)$ and $D_A(z)$ we can propagate it to the Fisher matrix for the
parameters using~\cite{Wang2010}
\begin{equation}
F_{\alpha\alpha'}=\sum_{ij}\frac{\partial p_i}{\partial
  q_{\alpha}}F_{ij}\frac{\partial p_j}{\partial q_{\alpha'}}\,,
\end{equation}
where $p_i$ stand for the observables $H(z)$ or $D_A(z)$, and
$q_{\alpha}$ for the parameters under study. The Fisher matrix for all
the redshift range of the survey is given by the sum of the Fisher
matrices for each redshift bin. The inverse of the resulting Fisher
matrix gives us the uncertainties and correlations of all the parameters
studied in the forecast.

\subsection{Euclid spectroscopic survey}
In this work we focus our forecast on the typical results expected for
the Euclid\,\footnote{http://www.euclid-ec.org} space mission: an ESA
medium class astronomy and astrophysics space mission which aims at
understanding the accelerated expansion of the Universe and the nature
of dark energy.

For simplicity, we restrict ourselves to the galaxy clustering probe
of the Euclid mission. In order to adapt our forecast we only need five
parameters, whose values are taken from the Euclid
redbook~\cite{Laureijs}: the minimum and maximum redshift,
$z_{\text{min}}=0.7$, $z_{\text{max}}=2.1$; the area, 15000 square
degrees; the number of galaxies, $50\times 10^6$; and the precision of
the redshift estimation, $\sigma_z/(1+z)\leq 0.1\%$. We adopt the bias
given in \cite{Amendola2013}: $b(z)=\sqrt{1+z}$. We split the redshift
range of the survey into bins of width 0.1 in redshift. Narrower bins
only marginally increase the precision while requiring more
computational time. Finally, the reference cosmology is the one obtained
in Sec.\,\ref{section:results} and it is summarized in the fifth column
of Table\,\ref{table1}.

In this work we limit ourselves to the
spectroscopic survey on linear scales. We have checked that the photometric survey only
marginally improves the constraints on the parameters, while including
weakly nonlinear scales noticeably improves these constraints. Combination with
the weak lensing probe would obviously lead to even better
constraints~\cite{Majerotto2012}. A quantitative evaluation of these improvements is left for
future work.

\subsection{Results}

\begin{figure*}[t]
\includegraphics[scale=0.45]{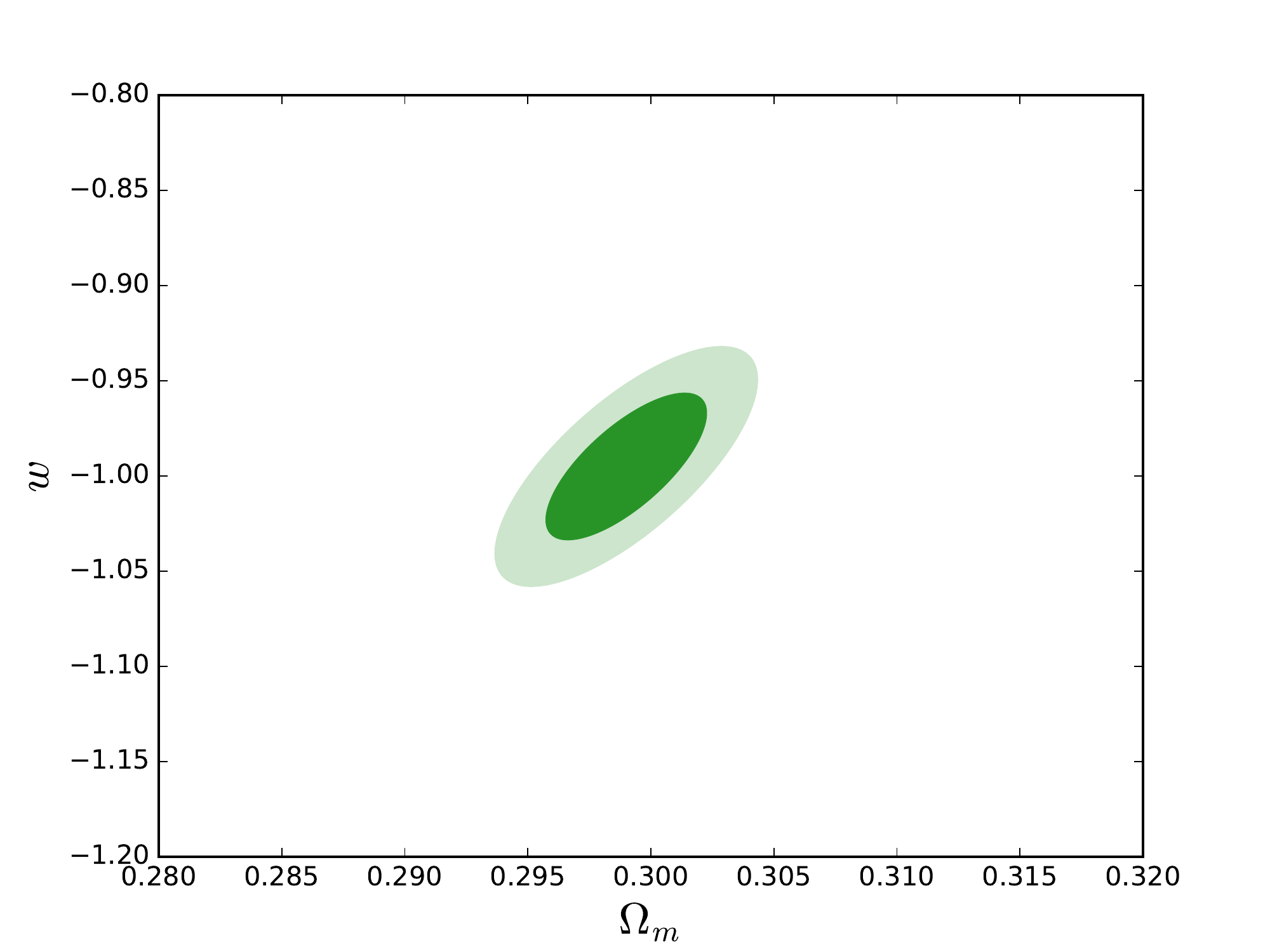}\includegraphics[scale=0.45]{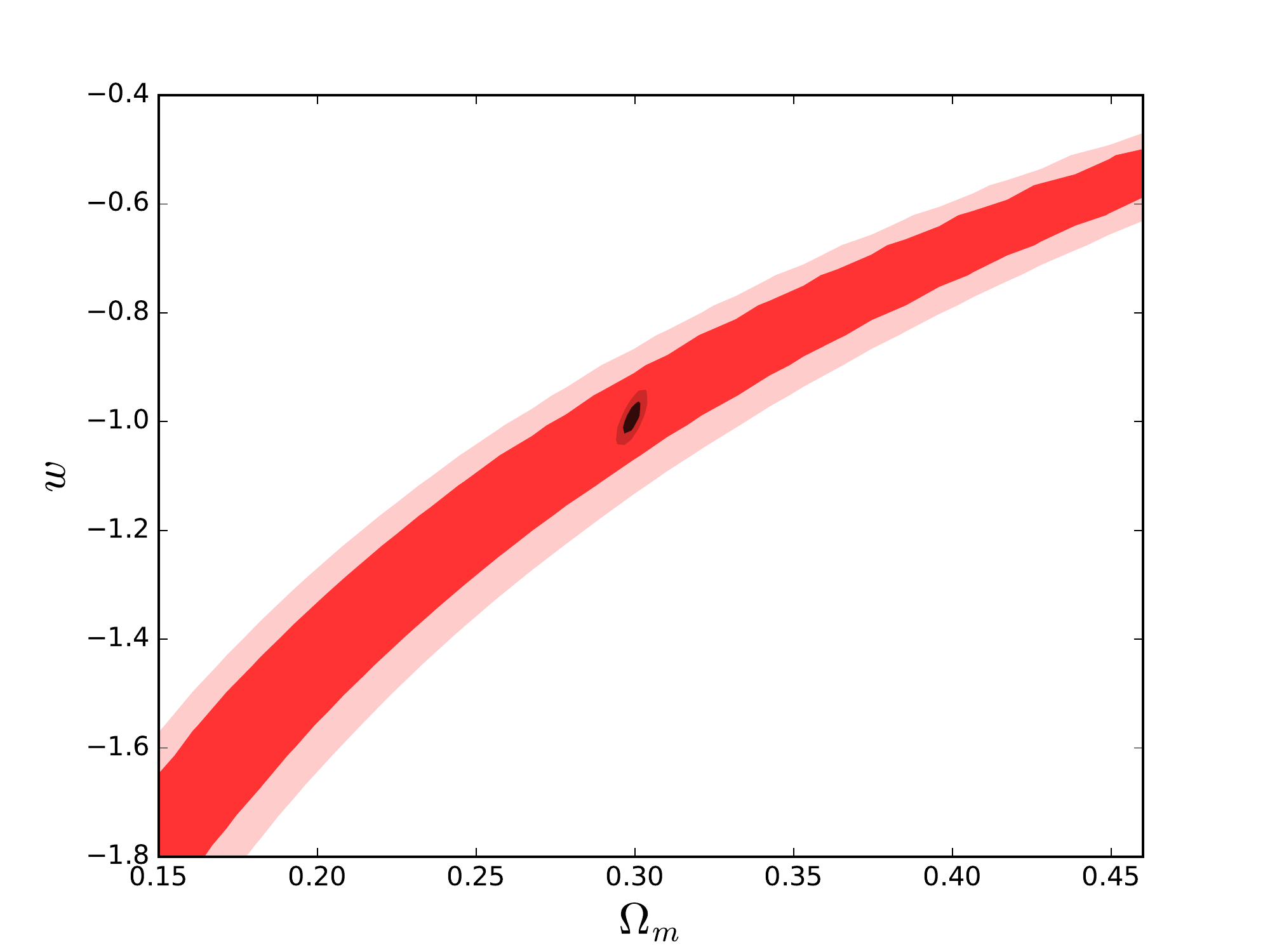}
\caption{Confidence contours at 68\% and 95\% ($\Delta \chi^2=2.30$
  and $\Delta \chi^2=6.17$, respectively) for the $\Omega_m$ and $w$
  cosmological parameters of the $w$CDM model. Left panel: contours
  obtained using the Euclid galaxy clustering forecast with
  $\omega_b=0.02262$ fixed. Right panel: the red (large) contours correspond
  to the CMB probe while the black (small) contours account for the
  combination of the CMB and the Euclid galaxy clustering forecast.}\label{fig4}
\end{figure*}

The results for the $w$CDM model are represented in
Fig. \ref{fig4}. In the left panel the 1$\sigma$ and 2$\sigma$
contours for the $\Omega_m$ and $w$ cosmological parameters are
computed using the Euclid galaxy power spectrum forecast and fixing
the reduced baryon density parameter to its reference value
$\omega_b=0.02257$. In the right panel the red contours correspond to
the CMB probe and the black contours stand for the combination of the
CMB and the forecast assuming a Gaussian likelihood for the
forecast. More precisely, when minimizing the $\chi^2$ function as
presented in (\ref{chisq}), we minimize the sum of the $\chi^2$
corresponding to the CMB plus a $\chi^2$ function associated to the
forecast where the covariance matrix is directly the one provided by
the forecast. We have marginalized over $H_0$ in all the figures. The
results of the forecast are the following constraints:
$\Omega_m=0.299\pm 0.022$ and $w=-0.995\pm 0.026$ (errors at
1$\sigma$ on one parameter), which are much better than the degeneracy
obtained with SNIa+BAO present-day data (Fig.\,\ref{fig1}, left
panel). Combination with the CMB gives $\Omega_m=0.2990\pm 0.0021$
and $w=-0.994\pm 0.022$ (errors at 1$\sigma$ on one parameter), which
are between a factor 2 and 6 better than SNIa+BAO+CMB present-day data
constraints (Fig.\,\ref{fig1}, right panel).

\begin{figure*}[t]
\includegraphics[scale=0.45]{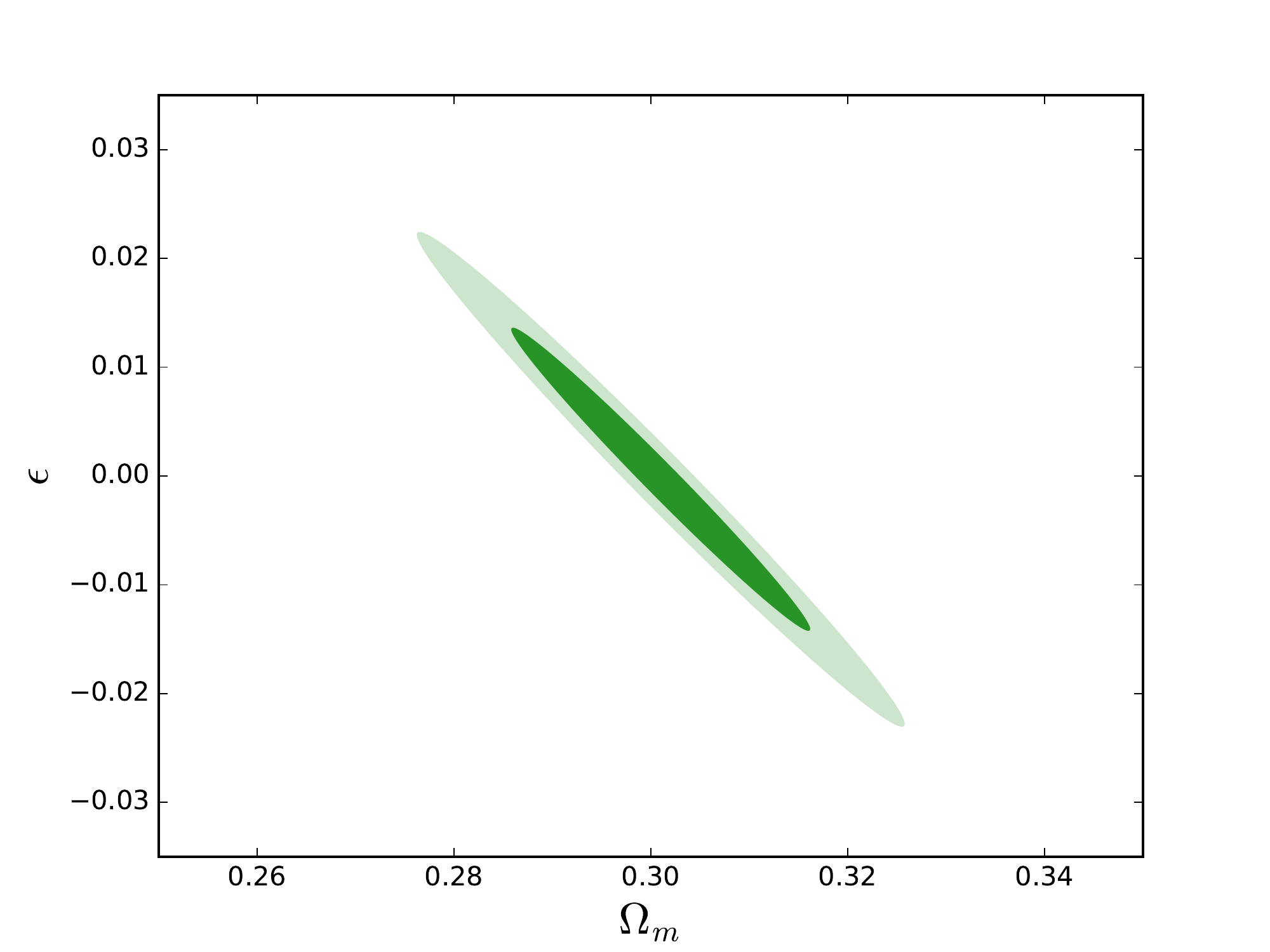}\includegraphics[scale=0.45]{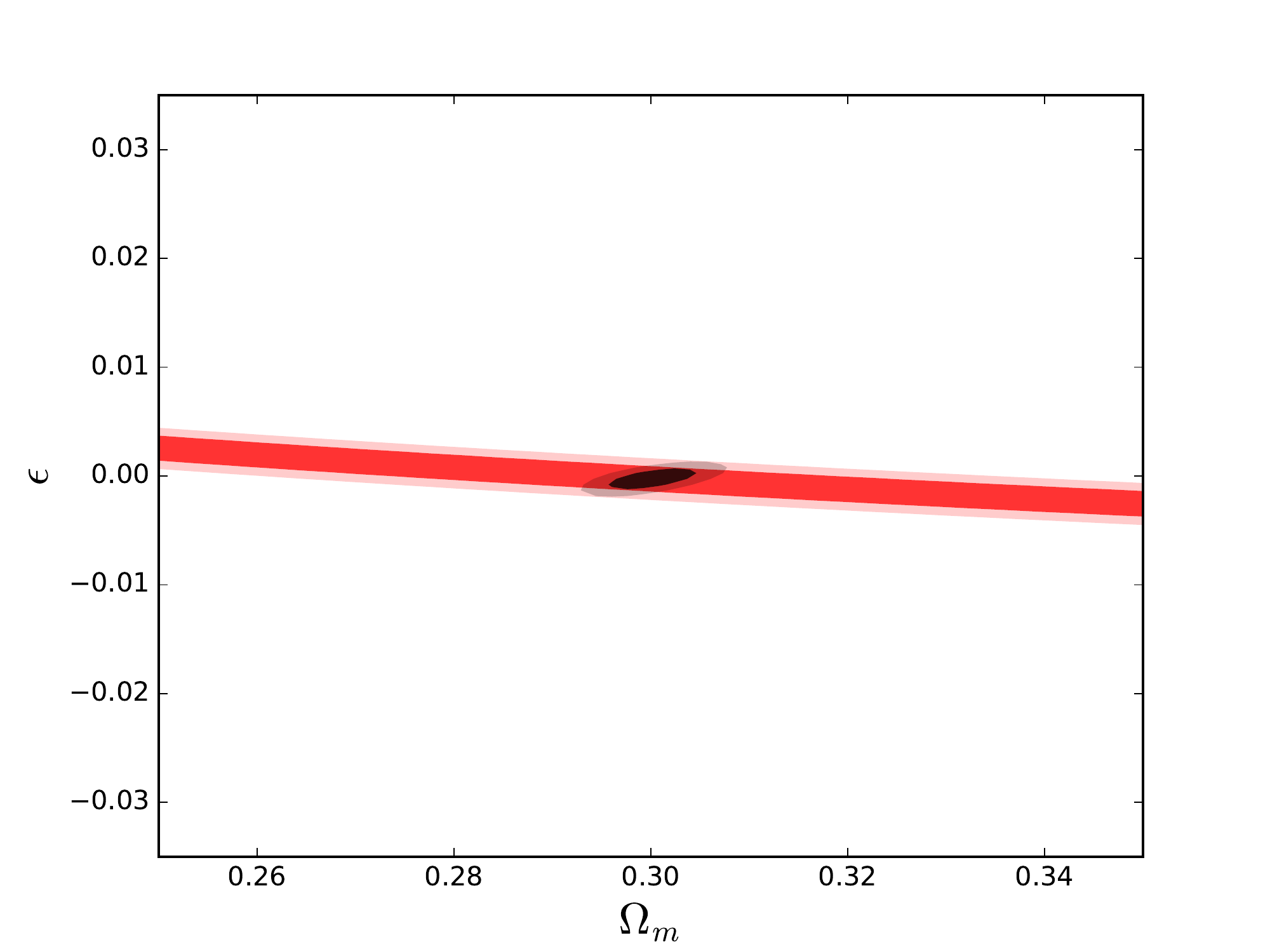}
\caption{Confidence contours at 68\% and 95\% ($\Delta \chi^2=2.30$
  and $\Delta \chi^2=6.17$, respectively) for the $\Omega_m$ and $\epsilon$
  cosmological parameters of the $\epsilon$CDM model. Left panel: contours
  obtained using the Euclid galaxy clustering forecast with
  $\omega_b=0.02262$ fixed. Right panel: the red (large) contours correspond
  to the CMB probe while the black (small) contours account for the
  combination of the CMB and the Euclid galaxy clustering forecast.}\label{fig5}
\end{figure*}

The 1$\sigma$ and 2$\sigma$ contours for the $\Omega_m$ and
$\epsilon$ cosmological parameters of the $\epsilon$CDM model are
represented in Fig.\,\ref{fig5}. As in Fig.\,\ref{fig4} the left panel
corresponds to the Euclid galaxy power spectrum forecast with fixed
baryon density, while the right panel corresponds to the CMB (red) and
the combination of the forecast and the CMB (black) contours. We have
also marginalized over $H_0$ in all the figures. The specific
constraints given by the forecast are $\Omega_m=0.301\pm 0.010$ and
$\epsilon=-0.0003\pm 0.0092$ (errors at 1$\sigma$ on one parameter),
which again have greatly improved compared to the degeneracy found with
SNIa+BAO present-day data (Fig.\,\ref{fig2}, left panel). Adding the
CMB we obtain the constraints, $\Omega_m=0.3001\pm 0.0030$ and
$\epsilon=-0.00024^{+0.00065}_{-0.00066}$ (errors at 1$\sigma$ on one
parameter), which are between a factor 2 and 5 better than present-day
SNIa+BAO+CMB constraints (Fig.\,\ref{fig2}, right panel).

Let us recall that all the results shown here are only for the galaxy clustering
probe restricted to the linear scales, so we can expect significantly better constraints from the full exploitation of the future Euclid survey data.

\begin{figure*}[t]
\includegraphics[scale=0.45]{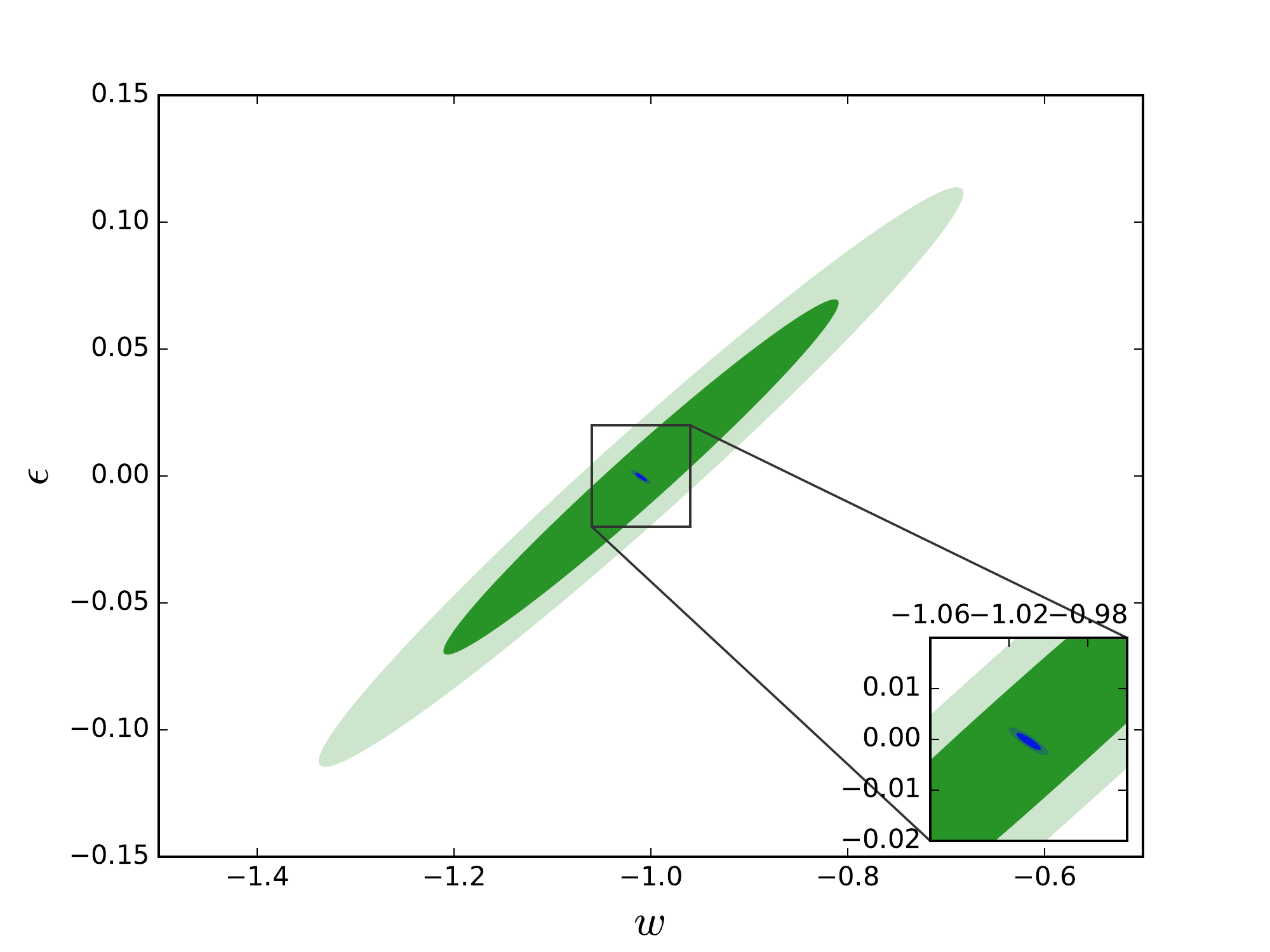}\includegraphics[scale=0.45]{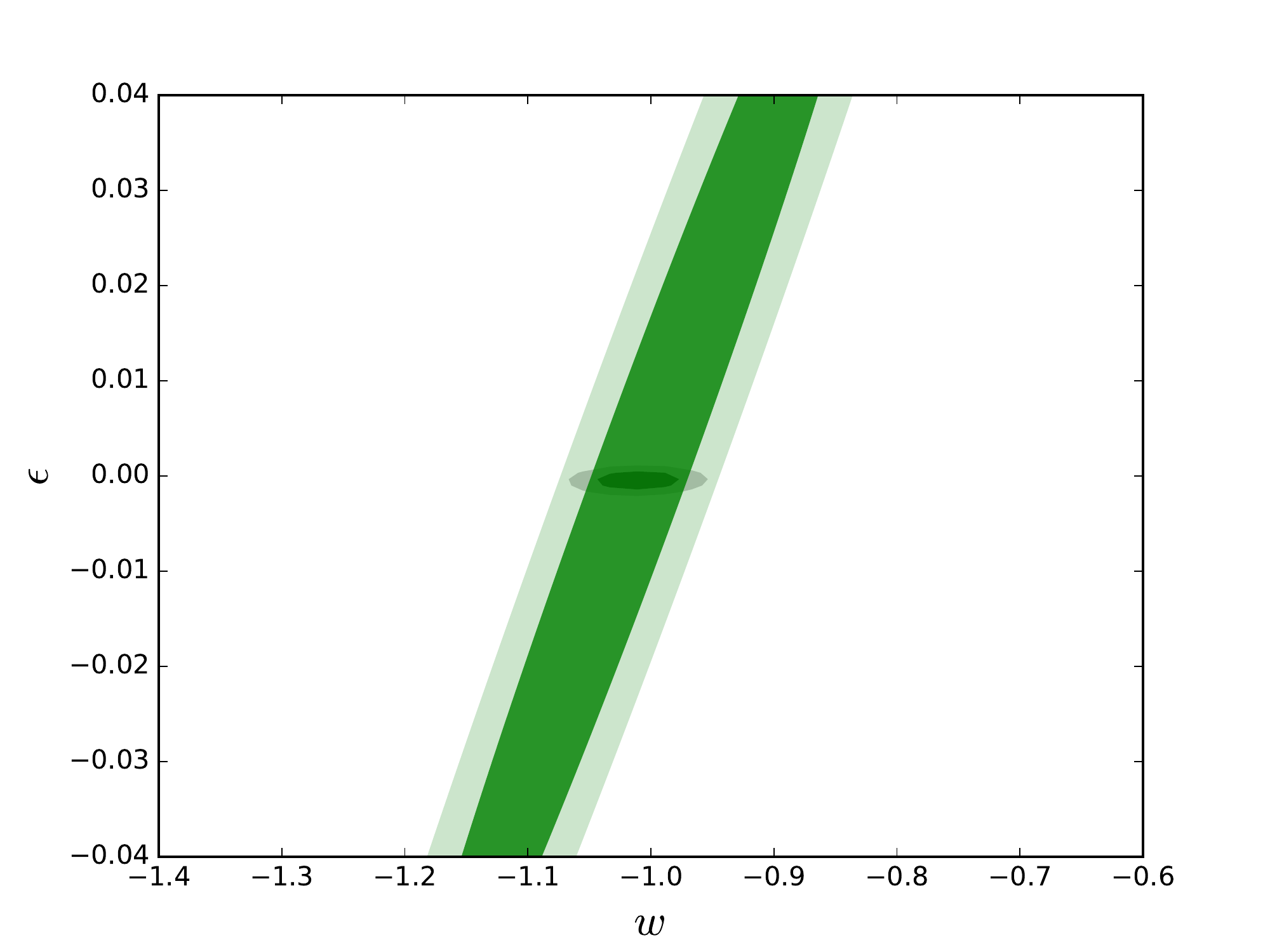}
\caption{Confidence contours at 68\% and 95\% ($\Delta \chi^2=2.30$
  and $\Delta \chi^2=6.17$, respectively) for the $\epsilon$ and $w$
  cosmological parameters of the $\epsilon w$CDM model. Left panel:
  contours obtained using the Euclid galaxy clustering forecast
  with $\omega_b=0.02262$ fixed. In the small box on the lower right corner, the equivalent contours when fixing $H_0=68.6$
  km\,s$^{-1}$\,Mpc$^{-1}$ and $\Omega_m=0.301$ are represented. Right
  panel: the green (large) contours correspond to the forecast while
  the black (small) contours
  account for the combination with the CMB, marginalizing over the baryon density.}\label{fig6}
\end{figure*}

Figure\,\ref{fig6} provides the 1$\sigma$ and 2$\sigma$ contours for
the $w$ and $\epsilon$ cosmological parameters of the $\epsilon w$CDM
model. The left panel corresponds to the forecast with fixed baryon
density, while the right panel shows, in addition, the combination of
the forecast with the CMB. We have marginalized over $H_0$ in both cases. The
specific constraints we have obtained using the forecast are
$w=-1.01\pm 0.13$ and $\epsilon=0.000\pm 0.046$ (errors at 1$\sigma$
on one parameter), which are much better than present-day SNIa+BAO
degeneracies that do not provide any significant constraint (see the
absence of constraints in the third column of Table\,\ref{table1}). This is a
remarkable result illustrating that Euclid can break this degeneracy in
the dark sector at low redshift.

It is worth noticing, apart from the better constraints expected from
the Euclid survey, that we still find\,\footnote{We have checked that
  changing (by a 20\% difference) the fixed value for the reduced
  baryon density parameter negligibly affects the $\epsilon-w$
  contours.} a significant correlation between the $\epsilon$ and $w$
cosmological parameters from the Euclid survey (Fig. \ref{fig6}, left
panel, green contour) illustrating the fact that the dark matter and
the dark energy sectors are not completely uncoupled and cannot be
constrained independently from each other. However, the sign of this
correlation may be somewhat surprising: if the total density were to
be constant we would expect $w$ and $\epsilon$ to be
anticorrelated. We have checked that this is indeed the case, when
all the other parameters are kept fixed (see the small box in Fig.\,\ref{fig6},
left panel). When marginalizing over $H_0$ and $\Omega_m$ the
correlation changes and leads to weak constraints on the dark energy
equation of state parameter, $w = -1.01\pm 0.13$ ($w = -1.01$ being
the fiducial value that corresponds to our best estimate in view of
present-day constraints), and on the equation of state of dark matter
$\epsilon = 0 \pm 0.046$.

Adding the CMB constraint to the forecast results in much more stringent
limits on the parameters describing the dark sector, $w=-1.010\pm
0.023$ and $\epsilon=-0.00045^{+0.00065}_{-0.00066}$ (errors at
1$\sigma$ on one parameter), which are similar to the obtained
constraints on the $w$CDM and the $\epsilon$CDM model parameters
(Figs.\,\ref{fig4} and \ref{fig5}, respectively). This fact
highlights again the strong role of the CMB in breaking degeneracies
thanks to the strong constraint on the dark matter equation of state
parameter.

\section{CONCLUSIONS}

We have investigated consequences on cosmological constraints relaxing
the pressureless nature of dark matter ($w_{DM}\neq0$). We restricted
ourselves to the simple case of constant equation of state parameter
for both dark sectors. Even if not fully theoretically motivated,
these simple models allow us to ascertain the maximum values that the
equation of state parameters are allowed to take~\cite{Kunz2016}.  We
have found that cosmological constraints from present-day SNIa and BAO
data are strongly degraded, revealing a complete degeneracy between the equations
of state of matter and dark energy.  The constraints are essentially
restored by the inclusion of CMB data thanks to its leverage. We have
then studied the anticipated accuracy from the Euclid redshift galaxy
survey. We have found that Euclid is expected to break the above degeneracy between
dark matter and dark energy, but the high accuracy on the dark energy
equation of state parameter is lost. Combining with the CMB
allows us to restore constraints at a similar level to the $w_{DM}=0$
forecast in the specific model we investigated.  We expect even better
performance from the full exploitation of the future Euclid survey
data, but the remaining correlation between dark matter and dark energy equation
of state parameter deserves further investigation.


\begin{table*}[t]
\caption{Cosmological parameter constraints for the different models
  and the different probes considered (Euclid GC stands for the galaxy clustering probe of the Euclid survey). The errors are given at the
  1$\sigma$ confidence level on one parameter ($\Delta \chi^2=1$). The $\Lambda$CDM model is
  included for comparison. The stars in some reduced baryon densities
  stand for fixed values. The dash in the $\epsilon w$CDM model using
SNIa+BAO data stands for the extreme degeneracies which do not allow
us to obtain significant constraints on the cosmological parameters.}\label{table1}
\begin{center}
\begin{tabular}{cc|c|c|c|c|}
\cline{3-6}
& & SNIa+BAO & Euclid GC & SNIa+BAO+CMB & Euclid GC + CMB\\
\cline{1-6}
\multicolumn{1}{|c}{\multirow{3}{*}{$\Lambda$CDM}} &
                                                     \multicolumn{1}{|c|}{$\Omega_m$} & $0.288^{+0.032}_{-0.031}$ & $0.2984^{+0.0015}_{-0.0015}$ & $0.2984^{+0.0096}_{-0.0092}$ &
                                                             $0.2984^{+0.0015}_{-0.0015}$\\

\multicolumn{1}{|c}{} & \multicolumn{1}{|c|}{$H_0$} & $67.6^{+2.7}_{-2.4}$ & $ 68.80^{+0.10}_{-0.10}$ & $68.80^{+0.75}_{-0.74}$ & $68.80^{+0.10}_{-0.10}$\\

\multicolumn{1}{|c}{} & \multicolumn{1}{|c|}{$\omega_b$} & $0.02262^*$ & $0.02257^*$ & $0.02257^{+0.00024}_{-0.00024}$ & $0.022574^{+0.000098}_{-0.000098}$\\

\hline

\multicolumn{1}{|c}{\multirow{4}{*}{$w$CDM}} & \multicolumn{1}{|c|}{$\Omega_m$} & $\leq 0.28$ & $0.299^{+0.022}_{-0.022}$ & $0.299^{+0.012}_{-0.011}$ &
                                                             $0.2990^{+0.0021}_{-0.0021}$\\

\multicolumn{1}{|c}{} & \multicolumn{1}{|c|}{$w$} & $-0.72^{+0.18}_{-0.25}$ & $-0.995^{+0.026}_{-0.026}$ & 
                                              $-0.995^{+0.052}_{-0.054}$
                                                           & $-0.994^{+0.022}_{-0.022}$\\

\multicolumn{1}{|c}{} & \multicolumn{1}{|c|}{$H_0$} & $53.0^{+13.3}_{-5.5}$ & $ 68.70^{+0.45}_{-0.45}$ & $68.7^{+1.3}_{-1.3}$ & $68.68^{+0.39}_{-0.40}$\\

\multicolumn{1}{|c}{} & \multicolumn{1}{|c|}{$\omega_b$} & $0.02262^*$ & $0.02259^*$ & $0.02259^{+0.00026}_{-0.00026}$ & $0.022581^{+0.000098}_{-0.000098}$\\

\hline

\multicolumn{1}{|c}{\multirow{4}{*}{$\epsilon$CDM}} & \multicolumn{1}{|c|}{$\Omega_m$} & $\geq 0.31$ & $0.301^{+0.010}_{-0.010}$ & $0.301^{+0.014}_{-0.013}$ &
                                                             $0.3001^{+0.0030}_{-0.0030}$\\

\multicolumn{1}{|c}{} & \multicolumn{1}{|c|}{$\epsilon$} & $-0.49^{+0.44}_{-0.20}$ & $-0.0003^{+0.0092}_{-0.0092}$ & 
                                              $-0.0003^{+0.0011}_{-0.0011}$
                                                           & $-0.00024^{+0.00065}_{-0.00066}$\\

\multicolumn{1}{|c}{} & \multicolumn{1}{|c|}{$H_0$} & $50.00^{+3.83}_{-0.90}$ & $ 68.60^{+0.27}_{-0.27}$ & $68.6^{+1.2}_{-1.2}$ & $68.62^{+0.12}_{-0.12}$\\

\multicolumn{1}{|c}{} & \multicolumn{1}{|c|}{$\omega_b$} & $0.02262^*$ & $0.02262^*$ & $0.02262^{+0.00029}_{-0.00029}$ & $0.02262^{+0.00029}_{-0.00029}$\\

\hline

\multicolumn{1}{|c}{\multirow{5}{*}{$\epsilon w$CDM}} & \multicolumn{1}{|c|}{$\Omega_m$} & & $0.301^{+0.041}_{-0.041}$ & $0.301^{+0.014}_{-0.013}$ &
                                                             $0.3011^{+0.0038}_{-0.0037}$\\

\multicolumn{1}{|c}{} & \multicolumn{1}{|c|}{$w$} &  & $-1.01^{+0.13}_{-0.13}$ & 
                                              $-1.010^{+0.075}_{-0.077}$
                                                           & $-1.010^{+0.023}_{-0.023}$\\

\multicolumn{1}{|c}{} & \multicolumn{1}{|c|}{$\epsilon$} & $-$ & $0.000^{+0.046}_{-0.046}$ & 
                                              $-0.0004^{+0.0016}_{-0.0016}$
                                                           & $-0.00045^{+0.00065}_{-0.00066}$\\

\multicolumn{1}{|c}{} & \multicolumn{1}{|c|}{$H_0$} &  & $ 68.6^{+1.0}_{-1.0}$ & $68.6^{+1.3}_{-1.3}$ & $68.60^{+0.44}_{-0.43}$\\

\multicolumn{1}{|c}{} & \multicolumn{1}{|c|}{$\omega_b$} & & $0.02262^*$ & $0.02262^{+0.00029}_{-0.00029}$ & $0.02262^{+0.00029}_{-0.00029}$\\

\hline

\end{tabular}
\end{center}
\end{table*}

\vspace{10pt}

\textbf{ACKNOWLEDGMENTS}

We thank St\'ephane Ili\'{c}  for useful comments.

\end{document}